\documentclass[10pt,a4paper]{article}
\usepackage{amsfonts}
\usepackage{amsbsy}
\usepackage{amssymb}
\usepackage{amsmath}
\usepackage{amsthm}
\newcommand{\titleart}[1]{\bigskip \begin{center}
\large \textbf{#1}\end{center}}
\vspace{-6mm}
\newcommand{\autorart}[1]{\begin{center} \large
\textsf{#1}\end{center} \vspace{-6mm} }
\newcommand{\coord}[1]{\begin{center} \small
\textit{#1}\end{center} \vspace{-6mm} }
\newcommand{\mail}[1]{\begin{center} \small
\textit{#1}\end{center} \normalsize}
\begin{document}
\titleart{Algorithmic Techniques for Several Optimization Problems Regarding Distributed Systems with Tree Topologies}
\autorart{Mugurel Ionu\c t Andreica \index{Mugurel Ionu\c t Andreica}}
\coord{Politehnica University of Bucharest, Computer Science Department, Romania}
\mail{mugurel.andreica@cs.pub.ro}
\medskip
As the development of distributed systems progresses, more and more challenges arise and the need for developing optimized systems and for optimizing existing systems from multiple perspectives becomes more stringent. In this paper I present novel algorithmic techniques for solving several optimization problems regarding distributed systems with tree topologies. I address topics like: reliability improvement, partitioning, coloring, content delivery, optimal matchings, as well as some tree counting aspects. Some of the presented techniques are only of theoretical interest, while others can be used in practical settings.
\medskip

\section{Introduction}
Distributed systems are being increasingly developed and deployed all around the world, because they present efficient solutions to many practical problems. However, as their development progresses, many problems related to scalability, fault tolerance, stability, efficient resource usage and many other topics need to be solved. Developing efficient distributed systems is not an easy task, because many system parameters need to be fine tuned and optimized. Because of this, optimization techniques are required for designing efficient distributed systems or improving the performance of existing, already deployed ones. In this paper I present several novel algorithmic techniques for some optimization problems regarding distributed systems with a tree topology.

Trees are some of the simplest non-trivial topologies which appear in real-life situations. Many of the existing networks have a hierarchical structure (a tree or tree-like graph), with user devices at the edge of the network and router backbones at its core. Some peer-to-peer systems used for content retrieval and indexing have a tree structure. Multicast content is usually delivered using multicast trees. Furthermore, many graph topologies can be reduced to tree topologies, by choosing a spanning tree or by covering the graph's edges with edge disjoint spanning trees [1]. In a tree, there exists a unique path between any two nodes. Thus, the network is quite fragile. The fragility is compensated by the simplicity of the topology, which makes many decisions become easier.

This paper is structured as follows. Section 2 defines the main notations which are used in the rest of the paper. In Section 3 I consider the minimum weight cycle completion problem in trees. In Section 4 I discuss two tree partitioning problems and in Section 5 I consider two content delivery optimization problems. In Section 6 I solve several optimal matching problems in trees and powers of trees and in Section 7 I analyze the {\it first fit online coloring} heuristic, applied to trees. In Section 8 I consider three other optimization and tree counting problems. In Section 9 I discuss related work and in Section 10 I conclude and present future work.

\section{Notations}

A tree is an undirected, connected, acyclic graph. A tree may be rooted, in which case a special vertex $r$ will be called its root. Even if the tree is unrooted, we may choose to root it at some vertex. In a rooted tree, we define $parent(i)$ as the parent of vertex $i$ and $ns(i)$ as the number of sons of vertex $i$. For a leaf vertex $i$, $ns(i)=0$ and for the root $r$, $parent(r)$ is undefined. The sons of a vertex $i$ are denoted by $s(i,j)$ $(1\le j\le ns(i))$. A vertex $j$ is a {\it descendant} of vertex $i$ if $(parent(j)=i)$ or $parent(j)$ is also a descendant of vertex $i$. We denote by $T(i)$ the subtree rooted at vertex $i$, i.e. the part of the tree composed of vertex $i$ and all of its descendants (together with the edges connecting them). In the paper, the terms {\it node} and {\it vertex} will be used with the same meaning.

A matching $M$ of a graph $G$ is a set of edges of the graph, such that any two edges in the set have distinct endpoints (vertices). A maximum matching is a matching with maximum cardinality (maximum number of edges).

\section{Minimum Weight Cycle Completion of a Tree}

We consider a tree network with $n$ vertices. For $m$ pairs of vertices $(i,j)$ which are not adjacent in the tree, we are given a weight $w(i,j)$ (we can consider $w(i,j)=+\infty$ for the other pairs of vertices). We want to connect some of these $m$ pairs (i.e. add extra edges to the tree), such that, in the end, every vertex of the tree belongs to exactly one cycle. The objective consists of minimizing the total weight of the edges added to the tree. For the unweighted case $(w(i,j)=1)$ and when we can connect any pair of vertices which is not connected by a tree edge, there exists the following simple greedy algorithm [3]. We select an arbitrary root vertex $r$ and then traverse the tree bottom-up (from the leaves towards the root). For each vertex $i$ we will compute a value $l(i)$, representing the largest number of vertices on a path $P(i)$ starting at $i$ and continuing in $T(i)$, such that every vertex $j \in (T(i) \setminus P(i))$ belongs to exactly one cycle and the vertices in $P(i)$ are the only ones who do not belong to a cycle. We denote by $e(i)$ the second endpoint of the path (the first one being vertex $i$). For a leaf vertex $i$, we have $l(i)=1$ and $e(i)=i$. For a non-leaf vertex $i$, we first remove from its list of sons the sons $s(i,j)$ with $l(s(i,j))=0$, update $ns(i)$ and renumber the other sons starting from $1$. If $i$ remains with only one son, we set $l(i)=l(s(i,1))+1$ and $e(i)=e(s(i,1))$. If $i$ remains with $ns(i)>1$ sons, we will sort them according to the values $l(s(i,j))$, such that $l(s(i,1)) \le l(s(i,2)) \le \ldots \le l(s(i,ns(i)))$. We will connect by an edge the vertices $e(s(i,1))$ and $e(s(i,2))$. This way, every vertex on the paths $P(s(i,1))$ and $P(s(i,2))$, plus the vertex $i$, belong to exactly one cycle. For the other sons $s(i,j)$ $(3\le j\le ns(i))$, we will have to connect $s(i,j)$ to $e(s(i,j))$. This will only be possible if $l(s(i,j))\ge 3$; otherwise, the tree admits no solution. Afterwards, we set $l(i)=0$. If the root $r$ has only one son, then we must have $l(r)\ge 3$, such that we can connect $r$ to $e(r)$.

For the general case, I will describe a dynamic programming algorithm (as the greedy algorithm cannot be extended to this case). We will again root the tree at an arbitrary vertex $r$, thus defining parent-son relationships. For each vertex $i$, we will compute two values: $wA(i)$=the minimum total weight of a subset of edges added to the tree such that every vertex in $T(i)$ belongs to exactly one cycle, and $wB(i)$=the minimum total weight of a subset of edges added to the tree such that every vertex in $(T(i) \setminus \{i\})$ belongs to exactly one cycle (and vertex $i$ belongs to no cycle). We will compute the values from the leaves towards the root. For a leaf vertex $i$, we have $wA(i)=+\infty$ and $wB(i)=0$. For a non-leaf vertex $i$, we have: $wB(i)=\sum_{j=1}^{ns(i)}wA(s(i,j))$. In order to compute $wA(i)$ we will first traverse $T(i)$ and for each vertex $j$, we will compute $wAsum(i,j)$=the sum of all the $wA(p)$ values, where $p$ is a son of a vertex $q$ which is located on the path from $i$ to $j$ $(P(i \ldots j))$ and $p$ does not belong to $P(i \ldots j)$. We have $wAsum(i,i)=wB(i)$ and for the other vertices $j$ we have $wAsum(i,j)=wAsum(i,parent(j))-wA(j)+wB(j)$. Now we will try to add an edge, such that it closes a cycle in the tree which contains vertex $i$. We will first try to add edges of the form $(i,j)$, where $j$ is a descendant of $i$ (but not a son of $i$, of course) - these will be called {\it type 1} edges. Adding such an edge $(i,j)$ provides a candidate value $wcand(i,i,j)$ for $wA(i)$: $wcand(i,i,j)=wAsum(i,j)+w(i,j)$. We will then consider edges of the form $(p,q)$ ($p\neq i$ and $q\neq i$), where the lowest common ancestor of $p$ and $q$ $(LCA(p,q))$ is vertex $i$ - these will be called {\it type 2} edges (we consider every pair of distinct sons $s(i,a)$ and $s(i,b)$, and for each such pair we consider every pair of vertices $p \in T(s(i,a))$ and $q \in T(s(i,b))$ and verify if the edge $(p,q)$ can be added to the tree). Adding such an edge $(p,q)$ provides a candidate value $wcand(i,p,q)$ for $wA(i)$: $wcand(i,p,q)$=$wAsum(i,p)$+$wAsum(i,q)$-$wB(i)$+$w(p,q)$. $wA(i)$ will be equal to the minimum of the candidate values $wcand(i,*,*)$ (or to $+\infty$ if no candidate value exists). We can implement the algorithm in $O(n^2)$ time, which is optimal in a sense, because $m\le (n\cdot (n-1)/2-n+1)$, which is $O(n^2)$. $wA(r)$ is the answer to our problem and we can find the actual edges to add to the tree by tracing back the way the $wA(*)$ and $wB(*)$ values were computed.

However, when the number $m$ of edges which can be added to the tree is significantly smaller, we can improve the time complexity to $O((n+m)\cdot log(n))$. We will compute for each of the $m$ edges $(i,j)$ the lowest common ancestor of the vertices $i$ and $j$ $(LCA(i,j))$ in the rooted tree. This can be achieved by preprocessing the tree in $O(n)$ time and then answering each LCA query in $O(1)$ time [2]. If $LCA(i,j)=k$, then we will add the edge $(i,j)$ to a list $Ledge(k)$. Then, for each non-leaf vertex $i$, we will traverse the edges in $Ledge(k)$. For each edge $(p,q)$ we can easily determine if it is of type $1$ ($i=p$ or $i=q$) or of type $2$ and use the corresponding equation. However, we need the values $wAsum(i,p)$ and $wAsum(i,q)$. Instead of recomputing these values from scratch, we will update them incrementally. It is obvious that $wAsum(parent(i),p)$=$wAsum(i,p)$+$wB(parent(i))$-$wA(i)$. We will preprocess the tree, by assigning to each vertex $i$ its DFS number $DFSnum(i)$ ($DFSnum(i)$=$j$ if vertex $i$ was the $j^{th}$ distinct vertex visited during a DFS traversal of the tree which started at the root). Then, for each vertex $i$, we compute $DFSmax(i)$=the maximum DFS number of a vertex in its subtree. For a leaf node $i$, we have $DFSmax(i)=DFSnum(i)$. For a non-leaf vertex $i$, $DFSmax(i)$=$max\{DFSnum(i),$ $DFSmax(s(i,1)),$ \ldots, $DFSmax(s(i,ns(i)))\}$. We will maintain a segment tree, using the algorithmic framework from [15]. The operations we will use are range addition update and point query. Initially, each leaf $i$ $(1 \le i \le n)$ has a value $v(i)=0$. Before computing $wA(i)$ for a vertex $i$, we set the value of leaf $DFSnum(i)$ in the segment tree to $wB(i)$. Then, for each son $s(i,j)$, we add the value $(wB(i)-wA(s(i,j))$ to the interval $[DFSnum(s(i,j)), DFSmax(s(i,j))]$ (range update). We can obtain $wAsum(i,p)$ for any vertex $p \in T(i)$ by querying the value of the cell $DFSnum(p)$ in the segment tree: we start from the (current) value of the leaf $DFSnum(p)$ and add the update aggregates {\it uagg} stored at every anestor node of the leaf in the segment tree. Queries and updates take $O(log(n))$ time each.

If the objective is to minimize the largest weight $W_{max}$ of an edge added to the tree, we can binary search $W_{max}$ and perform the following feasibility test on the values $W_{cand}$ chosen by the binary search: we consider only the "extra" edges $(i,j)$ with $w(i,j)\le W_{cand}$ and run the algorithm described above for these edges; if $wA(r)\neq +\infty$, then $W_{cand}$ is feasible.

\section{Tree Partitioning Techniques}

\subsection{Tree Partitioning with Lower and Upper Size Bounds}

Given a tree with $n$ vertices, we want to partition the tree into several parts, such that the number of vertices in each part is at least $Q$ and at most $k\cdot Q$ $(k \ge 1)$. Each part $P$ must have a representative vertex $u$, which does not necessarily belong to $P$. However, $(P \cup \{u\})$ must form a connected subtree. I will present an algorithm which works for $k\ge 3$. We root the tree at any vertex $r$, traverse the tree bottom-up and compute the parts in a greedy manner. For each vertex $i$ we compute $w(i)$=the size of a connected component $C(i)$ in $T(i)$, such that vertex $i \in C(i)$ , $|C(i)|<Q$, and all the vertices in $(T(i) \setminus C(i))$ were split into parts satisfying the specified properties. For a leaf vertex $i$, $w(i)=1$ and $C(i)=\{i\}$. For a non-leaf vertex $i$, we traverse its sons (in any order) and maintain a counter $ws(i)$=the sum of the $w(s(i,j))$ values of the sons traversed so far. If $ws(i)$ exceeds $Q-1$ after considering the son $s(i,j)$, we form a new part from the connected components $C(s(i,last\_son+1)), \ldots, C(s(i,j))$ and assign vertex $i$ as its representative. Then, we reset $ws(i)$ to $0$. $(last\_son<j)$ is the previous son where $ws(i)$ was reset to $0$ (or $0$, if $ws(i)$ was never reset to $0$).

After considering every son of vertex $i$, we set $w(i)=ws(i)+1$ and the component $C(i)$ is formed from the components $C(s(i,j))$ which were not used for forming a new part, plus vertex $i$. If $ws(i)+1=Q$, then we form a new part from the component $C(i)$ and set $w(i)=0$ and $C(i)=\{\}$. During the algorithm, the maximum size of any part formed is $2\cdot Q-2$. At the end of the algorithm, we may have that $w(r)>0$. In this case, the vertices in $C(r)$ were not assigned to any part. However, at least one vertex from $C(r)$ is adjacent to a vertex assigned to some part $P$. Then, we can extend that part $P$ in order to contain the vertices in $C(r)$. This way, the maximum size of a part becomes $3\cdot Q-3$. The pseudocode of the first part of the algorithm is presented below. In order to compute the parts, we maintain for each vertex $i$ a value $part(i)$, which is $0$, initially ($0$ means that the vertex was not assigned to any part). In order to assign distinct part numbers, we will maintain a global counter $part\_number$, whose initial value is $0$. The first part of the algorithm has linear time complexity ($O(n)$). The second part (adding $C(r)$ to an already existing part) can also be performed in linear time, by searching for an edge $(p,q)$, such that $part(p)=0$ and $part(q)>0$ (there are only $n-1=O(n)$ edges in a tree).\vspace{2mm}

\noindent \hspace{2mm}
\underline{\bf LowerUpperBoundTreePartitioning(Q, i):} {\it

\noindent \hspace{2mm}
{\bf if} (ns(i)=0) {\bf then} w(i)=1 {\bf else}

\noindent \hspace{4mm}
  ws(i)=last\_son=0

\noindent \hspace{4mm}
  {\bf for} j=1 {\bf to} ns(i) {\bf do} // j=1,2,\ldots,ns(i)

\noindent \hspace{6mm}
    {\bf LowerUpperBoundTreePartitioning}(Q, s(i,j))

\noindent \hspace{6mm}
    ws(i)=ws(i)+w(s(i,j))

\noindent \hspace{6mm}
    {\bf if} $(ws(i)\ge Q)$ {\bf then}

\noindent \hspace{8mm}
      part\_number=part\_number + 1; last\_son=j; ws(i)=0

\noindent \hspace{8mm}
      {\bf for} k=last\_son+1 {\bf to} j {\bf do} {\bf AssignPartNumber}(s(i,k), part\_number)

\noindent \hspace{4mm}
  w(i)=ws(i)+1

\noindent \hspace{4mm}
  {\bf if} $(w(i) \ge Q)$ {\bf then}

\noindent \hspace{6mm}
    part\_number=part\_number + 1; w(i)=0

\noindent \hspace{6mm}
    {\bf AssignPartNumber}(i, part\_number)
} \vspace{2mm}

{\it
\noindent \hspace{2mm}
\underline{\bf AssignPartNumber(i, part\_number):}

\noindent \hspace{2mm}
{\bf if} $(part(i)\neq 0)$ {\bf then return()}

\noindent \hspace{2mm}
part(i)=part\_number

\noindent \hspace{2mm}
{\bf for} j=1 {\bf to} ns(i) {\bf do} {\bf AssignPartNumber}(s(i,j), part\_number)
}

\subsection{Connected Tree Partitioning}

I will now present an efficient algorithm for identifying $k$ connected parts of given sizes in a tree (if possible), subject to minimizing the total cost. Thus, given a tree with $n$ vertices, we want to find $k$ vertex-disjoint components (called parts), such that the $i^{th}$ part $(1\le i\le k)$ has $sz(i)$ vertices $(sz(1)+sz(2)+\ldots +sz(k)\le n$ and $sz(i)\le sz(i+1)$ for $1\le i\le k-1$). Each tree edge $(i,j)$ has a cost $ce(i,j)$ and each tree vertex $i$ has a cost $cv(i)$. We want to minimize the sum of the costs of the vertices and edges which do not belong to any part. An edge $(i,j)$ belongs to a part $p$ if both vertices $i$ and $j$ belong to part $p$.

In order to obtain $k$ connected components of the given sizes we need to keep $Q-k$ edges of the tree and remove the others, where $Q=sz(1)+\ldots +sz(n)$. We could try all the {\it ($(n-1)$ choose $(Q-k)$)} possibilities of choosing $Q-k$ edges out of the $n-1$ edges of the tree. For each possibility, we obtain $k'=n-Q+k$ connected components with sizes $sz'(1)\le sz'(2)\le \ldots \le sz'(k')$; in case of several components with equal sizes, we sort them in increasing order of the total cost of the vertices in them. Then, we must have $sz(j)=sz'(k'-k+j)$ and the total cost of the possibility is the sum of the costs of the removed edges plus the sum of the costs of the vertices in the components $1,2,\ldots,k'-k$ (which should have only one vertex each, if the size conditions hold). However, this approach is quite inefficient in most cases. I will present an algorithm with time complexity $O(n^3\cdot  3^k)$. We root the tree at an arbitrary vertex $r$. Then, we compute a table $Cmin(i,j,S)$=the minimum cost of obtaining from $T(i)$ the parts with indices in the set $S$ and, besides them, we are left with a connected component consisting of $j$ vertices which includes vertex $i$ and, possibly, several vertices which are ignored (if $j=0$, then every vertex in $T(i)$ is assigned to one of the parts in $S$ or is ignored). We compute this table bottom-up: \vspace{2mm}

\noindent \hspace{2mm}
\underline{\bf ConnectedTreePartitioning(i):} {\it

\noindent \hspace{2mm}
{\bf for each}  $S \subseteq \{1,2,\ldots,k\}$ {\bf do} {\bf for} j=0 {\bf to} n {\bf do} Cmin(i,j,S)=$+\infty$

\noindent \hspace{2mm}
Cmin(i, 1, \{\})=0; Cmin(i, 0, \{\})=cv(i)

\noindent \hspace{2mm}
{\bf for} x=1 {\bf to} ns(i) {\bf do}

\noindent \hspace{4mm}
  {\bf ConnectedTreePartitioning}(s(i,x))

\noindent \hspace{4mm}
  {\bf for each} $S \subseteq \{1,2,\ldots,k\}$ {\bf do} {\bf for} j=0 {\bf to} n {\bf do}

\noindent \hspace{6mm}
      Caux(i,j,S)=Cmin(i,j,S); Cmin(i,j,S)=$+\infty$

\noindent \hspace{4mm}
  {\bf for each} $S \subseteq \{1,2,\ldots,k\}$ {\bf do} {\bf for} j=0 {\bf to} n {\bf do}

\noindent \hspace{6mm}
        {\bf for each} $W \subseteq S$ {\bf do} {\bf for} q=0 {\bf to} qlimit(j) {\bf do}

\noindent \hspace{8mm}
Cmin(i,j,S)=min\{Cmin(i,j,S), Caux(i,j-q,$S\setminus W$) + extra\_cost(i,s(i,x),q) + Cmin(s(i,x),q,W)\} 

\noindent \hspace{2mm}
{\bf for each} $S \subseteq \{1,2,\ldots,k\}$ {\bf do}

\noindent \hspace{4mm}
  {\bf for} j=0 {\bf to} n {\bf do} {\bf if} ($Cmin(i, j, S)<+\infty$) {\bf then}

\noindent \hspace{6mm}
      {\bf for} q=1 {\bf to} k {\bf do} {\bf if} ((j=sz(q)) {\bf and} ($q \notin S$)) {\bf then}

\noindent \hspace{8mm}
        Cmin(i,0,$S \cup \{q \}$)=min\{Cmin(i,j,S), Cmin(i,0,$S \cup \{q \}$)\}
} \vspace{2mm}

We define {\it extra\_cost(i, son\_x\_i, q)=if $(q > 0)$ then return(0) else return(ce(i, son\_x\_i))} and {\it qlimit(j)=max\{j-1,0\}}. The algorithm computes {\it Cmin(i,*,*)} from the values of vertex $i$'s sons, using the principles of tree knapsack. The total amount of computations for each vertex is $O(ns(i)\cdot  3^k \cdot  n^2)$. Summing over all the vertices, we obtain $O(n^3 \cdot  3^k)$. The minimum total cost is $Cmin(r, 0, \{1,2,\ldots,k\})$ (if this value is $+\infty$, then we cannot obtain $k$ parts with the given sizes). In order to find the actual parts, we need to trace back the way the $Cmin(*,*,*)$ values were computed, which is a standard procedure. When the sum of the sizes of the $k$ parts is $n$, then every vertex belongs to one part.

\section{Content Delivery Optimization Problems}

\subsection{Minimum Number of Unicast Streams}

We consider a directed acyclic graph $G$ with $n$ vertices and $m$ edges. Every directed edge $(u,v)$ has a lower bound $lbe_{G}(u,v)$, an upper bound $ube_{G}(u,v)$ and a cost $ce_{G}(u,v)$. Every vertex $u$ has a lower bound $lbv_{G}(u)$, an upper bound $ubv_{G}(u)$ and a cost $cv_{G}(u)$. We need to determine the minimum number of unicast communication streams $p$ and a path for each of the $p$ streams, such that the number of stream paths $npe(u,v)$ containing an edge $(u,v)$ satisfies $lbe_{G}(u,v)\le npe(u,v)\le ube_{G}(u,v)$ and the number of paths $npv(u)$ containing a vertex $u$ satisfies $lbv_{G}(u)\le npv(u)\le ubv_{G}(u)$. Each vertex $u$ can be a {\it source} node, a {\it destination} node, both or none. A stream's path may start at any source node and finish at any destination node. Moreover, for the number of streams $p$, we want to compute the paths such that the sum $S$ over all the values $(npe(u,v)-lbe_{G}(u,v))$$\cdot ce_{G}(u,v)$ and $(npv(u)-lbv_{G}(u))$$\cdot cv_{G}(u)$ is minimum.

Particular cases of this problem have been studied previously. When {\it $lbv_{G}(u)$ =1} and $ubv_{G}(u)=1$ for every vertex $u$, $lbe_{G}(u,v)=0$ and $ube_{G}(u,v)=+\infty$ for every directed edge $(u,v)$, all the costs are $0$, and every vertex is a source and destination node, we obtain the {\it minimum path cover} problem in directed acyclic graphs, which is solved as follows [18]. Construct a bipartite graph $B$ with $n$ vertices $x_1,\ldots , x_n$ on the left side and $n$ vertices $y_1,\ldots ,y_n$ on the right side. We add an edge $(x_i,y_j)$ in $B$ if the directed edge $(i,j)$ appears in $G$. Then, we compute a maximum matching in $B$. If the cardinality of this matching is $C$, then we need $p=n-C$ streams. The paths are computed as follows. Having an edge $(x_i,y_j)$ in the maximum matching means that the edge $(i,j)$ in $G$ belongs to some stream's path. If two edges $(x_i,y_j)$ and $(x_j,y_k)$ in $B$ belong to the matching, then the edges $(i,j)$ and $(j,k)$ in $G$ belong to the path of the same stream. For non-zero costs, we compute a minimum (total) weight matching in $B$ (where every edge $(x_i,y_j)$ has a weight equal to $ce(i,j)$).

In order to solve the problem I mentioned, we will use a standard transformation and construct a new graph $G'$ where every vertex $u$ is represented by two vertices $u_{in}$ and $u_{out}$. For every directed edge $(u,v)$ in $G$, we add an edge $(u_{out},v_{in})$ in $G'$, with the same cost and lower and upper bounds. We also add a directed edge from $u_{in}$ to $u_{out}$ in $G'$ (for every vertex $u$ in $G$), with cost $cv_{G}(u)$, lower bound $lbv_{G}(u)$ and upper bound $ubv_{G}(u)$. Then we add two special vertices $s$ (source) and $t$ (sink) to $G'$. For every source node $u$ in $G$, we add a directed edge $(s,u_{in})$ in $G'$, with lower bound and cost $0$ and upper bound $+\infty$. For every destination node $v$ in $G$, we add a directed edge $(v_{out},t)$, with lower bound and cost $0$ and upper bound $+\infty$. We also add the edges $(s,t)$ and $(t,s)$ with lower bound and cost $0$ and upper bound $+\infty$. The resulting graph $G'$ has costs, lower and upper bounds only on its edges and not on its vertices. In order to compute the minimum number of communication streams which satisfy the constraints imposed by $G$, it is enough to compute a (minimum cost) minimum feasible flow in $G'$, from $s$ to $t$. Decomposing the flow into unit-flow paths (in order to obtain the path of each communication stream) can then be done easily. We repeatedly perform a graph traversal (DFS or BFS) from $s$ to $t$ in $G'$, considering only directed edges with positive flow on them. From the traversal tree, by following the "parent" pointers, we can find a path $P$ from $s$ to $t$, containing only edges with positive flow. We compute the minimum flow $fP$ on any edge of $P$, transform $P$ into $fP$ unit paths and then decrease the flow on the edges in $P$ by $fP$. If we remove the first and last vertices on any unit path (i.e. $s$ and $t$), we obtain a path from a vertex $u_{in}$ to a vertex $v_{out}$, where $u$ is a source node in $G$ and $v$ is a destination node in $G$. We will use the algorithm presented in [18] for determining a feasible flow (not necessarily minimum) in a flow network with lower and upper bounds on its edges. We will denote this algorithm by $A(F,s,t)$ ($F$ is the flow network given as argument, $s$ is the source vertex and $t$ is the sink vertex). I will describe $A(F,s,t)$ briefly. We construct a new graph $F'$ from $F$, as follows. We maintain all the vertices and edges in $F$. For every directed edge $(u,v)$ in $F$, the directed edge $(u,v)$ in $F'$ has the same cost, lower bound $0$ and upper bound $(ube_{F}(u,v)-lbe_{F}(u,v))$. We add two extra vertices $s'$ and $t'$ and the following zero-cost directed edges: $(s',u)$ and $(u,t')$ for every vertex $u$ in $F$ (including $s$ and $t$). The lower bound of every edge will be $0$. The upper bound of a directed edge $(s',u)$ in $F'$ is equal to the sum of the lower bounds of the directed edges $(*,u)$ in $F$. The upper bound of every directed edge $(u,t')$ in $F'$ is equal to the sum of the lower bounds of the directed edges $(u,*)$ in $F$. The algorithm $A(F,s,t)$ computes a minimum cost maximum flow $g$ in the graph $F'$ (which, as stated, only has upper bounds); if all the costs are $0$, only a maximum flow is computed. If $g$ is equal to the sum of the upper bounds of the edges $(s',*)$ (or, equivalently, of the edges $(*,t')$), then a feasible flow from $s$ to $t$ exists in $F$: the flow on every directed edge $(u,v)$ in $F$ will be $lbe_{F}(u,v)$ plus the flow on the edge $(u,v)$ in $F'$.

We will first run the algorithm on $G'$ (i.e. call $A(G',s,t)$) in order to verify if a feasible flow exists). If no feasible flow exists, then the constraints cannot be satisfied by any number of streams. Otherwise, we construct a graph $G''$ from $G'$, by adding a new vertex $snew$ and a zero-cost directed edge $(snew,s)$ with lower bound $0$ and upper bound $x$. $snew$ will be the new source vertex and $x$ is a parameter which is used in order to limit the amount of flow entering the old source vertex $s$. We will now perform a binary search on $x$, between $0$ and $gmax$, where $gmax$ is the value of the feasible flow computed by calling $A(G',s,t)$. The feasibility test consists of verifying if there exists a feasible flow in the graph $G''$ (i.e. calling $A(G'',snew,t)$). The minimum value of $x$ for which a feasible flow exists in $G''$ is the value of the minimum feasible flow in $G'$, from $s$ to $t$. Obtaining the feasible flow in $G'$ from the feasible flow in $G''$ is trivial: for every directed edge $(u,v)$ in $G'$, we set its amount of flow to the flow of the same edge $(u,v)$ in $G''$. The time complexity of the presented algorithm is $O(MF(n,m)\cdot log(gmax))$, where $gmax$ is a good upper bound on the value of a feasible flow and $MF(n,m)$ is the best time complexity of a (minimum cost) maximum flow algorithm in a directed graph with $n$ vertices and $m$ edges.

\subsection{Degree-Constrained Minimum Spanning Tree}

In [13], the following problem was considered: given an undirected graph with $n$ verices and $m$ edges, where each edge $(i,j)$ has a weight $w(i,j)>0$, compute a spanning tree $MST$ of minimum total weight, such that a special vertex $r$ has degree exactly $k$ in $MST$. A solution was proposed, based on using a parameter $d$ and setting the cost of each edge $(r,j)$ adjacent to $r$, $c(r,j)=d+w(r,j)$; the cost of the other edges is equal to their weight. Parameter $d$ can range from $-\infty$ to $+\infty$. We denote by $MST(d)$=the minimum spanning tree using the cost functions defined previously. When $d=-\infty$, $MST(d)$ contains the maximum number of edges adjacent to $r$. For $d=+\infty$, $MST(d)$ contains the minimum number of edges adjacent to $r$. We define the function $ne(d)$=the number of edges adjacent to $r$ in $MST(d)$. $ne(d)$ is non-increasing on the interval $[-\infty,+\infty]$. We will binary search the smallest value $dopt$ of the parameter $d$ in the interval $[-\infty,+\infty]$, such that $ne(dopt)\le k$. We will finish the binary search when the length of the search interval is smaller than a small constant $\varepsilon>0$.

If $ne(dopt)=k$, then the edges in $MST(dopt)$ form the required minimum spanning tree. If $ne(dopt)<k$, then $ne(dopt-\varepsilon)>k$. We define $S(d)$=the set of edges adjacent to vertex $r$ in $MST(d)$. It is easy to prove that $S(dopt)$ is included in $S(dopt-\varepsilon)$. The required minimum spanning tree is constructed in the following manner. The edges adjacent to vertex $r$ will be the edges in $S(dopt)$, to which we add $(k-ne(dopt))$ arbitrary edges from the set $S(dopt-\varepsilon) \setminus S(dopt)$. Once these edges are fixed, we construct the following graph $G$: we set the cost of the chosen edges to $0$ and the cost of the other edges $(i,j)$ to $w(i,j)$. We now compute a minimum spanning tree $MST_{G}$ in $G$. The edges in $MST_{G}$ are the edges of the minimum spanning tree of the original graph, in which vertex $r$ has degree exactly $k$. The time complexity of this approach is $O(m\cdot log(m)\cdot log(DMAX))$, where $DMAX$ denotes the range over which we search the parameter $d$. When $m$ is not too large (i.e. $m$ is not of the order $O(n^2$)), this represents an improvement over the $O(n^2)$ solution given in [13].

\section{Matching Problems}

\subsection{Maximum Weight Matching in an Extended Tree}

Let's consider a rooted tree (with vertex $r$ as the root). Each vertex $i$ has a weight $w(i)$. We want to find a matching in the following graph $G$ (extended tree), having the same vertices as $T$ and an edge $(x,y)$ between two vertices $x$ and $y$, if: {\it (i)} $x$ and $y$ are adjacent in the tree; {\it (ii)} $x$ and $y$ have the same parent in the tree. The weight of an edge $(x,y)$ in $G$ is $|w(x)-w(y)|$. The weight of a matching is the sum of the weights of its edges. We are interested in a maximum weight matching in the graph $G$. For each vertex $i$, we sort its sons $s(i,1), \ldots, s(i,ns(i))$ in non-decreasing order of their weights, i.e. $w(s(i,1)) \le \ldots \le w(i,ns(i))$. We will compute for each vertex $i$ two values: $A(i)$=the maximum weight of a matching in $T(i)$ if vertex $i$ is the endpoint of an edge in the matching and $B(i)$=the maximum weight of a matching in $T(i)$ if vertex $i$ is not the endpoint of any edge in the matching. In order to compute these values, we will compute the following tables for every vertex $i$: $CA(i,j,k)$=the maximum weight of a matching in $T(i)$ if vertex $i$ is the endpoint of an edge in the matching and we only consider its sons $s(i,j), s(i,j+1), \ldots, s(i,k)$ (and their subtrees). Similarly, we have $CB(i,j,k)$, where vertex $i$ does not belong to any edge in the matching. The maximum weight of a matching is $max\{A(r), B(r)\}$. The actual matching can be computed easily, by tracing back the way the $A(i)$, $B(i)$, $CA(i,*,*)$ and $CB(i,*,*)$ values were computed. A recursive algorithm (called with $r$ as its argument) is given below. The time complexity is $O(ns(i)^2)$ for a vertex $i$ and, thus, $O(n^2)$ overall.\vspace{2mm}

\noindent \hspace{2mm}
\underline{\bf MaximumWeightMatching-ExtendedTree(i):} {\it

\noindent \hspace{2mm}
{\bf if} (ns(i)=0) {\bf then} A(i)=B(i)=0 {\bf else}

\noindent \hspace{4mm}
{\bf for} j=1 {\bf to} ns(i) {\bf do} {\bf MaximumWeightMatching-ExtendedTree}(s(i,j))

\noindent \hspace{4mm}
{\bf for} j=1 {\bf to} ns(i) {\bf do}

\noindent \hspace{6mm}
  CA(i, j, j - 1)= $-\infty$; $CA(i, j, j)=|w(i) - w(s(i,j))| + B(s(i,j))$

\noindent \hspace{6mm}
  CB(i, j, j - 1)= 0; CB(i, j, j)=max\{A(s(i,j)), B(s(i,j))\}

\noindent \hspace{4mm}
{\bf for} count=1 {\bf to} (ns(i)-1) {\bf do} {\bf for} j=1 {\bf to} (ns(i)-count) {\bf do}

\noindent \hspace{6mm}
    k = j + count

\noindent \hspace{6mm}
    CA(i,j,k)=max\{$|w(s(i,j))-w(s(i,k))|$ + B(s(i,j)) + B(s(i,k)) + CA(i, j + 1, k - 1), $|w(i)-w(s(i, j))|$ + B(s(i,j)) + CB(i, j+1, k), $|w(i)-w(s(i, k))|$ + B(s(i,k)) + CB(i, j, k-1), max\{A(s(i,j)), B(s(i,j))\} + CA(i, j+1, k), max\{ A(s(i,k)), B(s(i,k))\} + CA(i, j, k-1)\}

\noindent \hspace{6mm}
    CB(i,j,k)=max\{$|w(s(i,j))-w(s(i,k))|$ + B(s(i,j)) + B(s(i,k)) + CB(i, j + 1, k - 1), max\{A(s(i,j)), B(s(i,j))\} + CB(i, j+1, k), max\{A(s(i,k)), B(s(i,k))\} + CB(i, j, k-1)\}

\noindent \hspace{4mm}
A(i)=CA(i,1,ns(i)); B(i)=CB(i,1,ns(i)) }

\subsection{Maximum Matching in the Power of a Graph}

The $k^{th}$ power $G^k$ ($k\ge 2)$ of a graph $G$ is a graph with the same set of vertices as $G$, where there exists an edge $(x,y)$ between two vertices $x$ and $y$ if the distance between $x$ and $y$ in $G$ is at most $k$. The distance between two vertices $(x,y)$ in a graph is the minimum number of edges which need to be traversed in order to reach vertex $y$, starting from vertex $x$. A maximum matching in $G^k$ of a graph $G$ can be found by restricting our attention to a spanning tree $T$ of $G$. The following linear algorithm (called with $i=r$), using observations from [12], solves the problem (we consider that, initially, no vertex is matched): \vspace{2mm}

\noindent \hspace{2mm}
\underline{\bf MaximumMatchingGk(i):} {\it

\noindent \hspace{2mm}
{\bf if} (ns(i)=0) {\bf then return()} {\bf else}

\noindent \hspace{4mm}
last\_son=0

\noindent \hspace{4mm}
{\bf for} j=1 {\bf to} ns(i) {\bf do} // j=1,2,\ldots ,ns(i)

\noindent \hspace{6mm}
{\bf MaximumMatchingGk}(s(i,j))

\noindent \hspace{6mm}
  {\bf if} ({\bf not} matched(s(i,j)) {\bf then}

\noindent \hspace{8mm}
    {\bf if} (last\_son = 0) {\bf then} last\_son = s(i,j) {\bf else}

\noindent \hspace{10mm}
          {\bf add edge} (last\_son, s(i,j)) {\bf to the matching}

\noindent \hspace{10mm}
          matched(last\_son) = matched(s(i,j)) = true; last\_son = 0

\noindent \hspace{4mm}
{\bf if} ($last\_son > 0$) {\bf then}

\noindent \hspace{6mm}
    {\bf add edge} (i, last\_son) {\bf to the matching}

\noindent \hspace{6mm}
    matched(i) = matched(last\_son) = true
} \vspace{2mm}

\section{First Fit Online Tree Coloring}

A very intuitive algorithm for coloring a graph with $n$ vertices is the {\it first-fit online coloring heuristic}. We traverse the vertices in some order $v(1), v(2), \ldots, v(n)$. We assign color $1$ to $v(1)$ and for $i=2,\ldots ,n$, we assign to $v(i)$ the minimum color $c(i)\ge 1$ which was not assigned to any of its neighbours $v(j)$ ($j<i$).

A tree is {\it 2-colorable}: we root the tree at any vertex $r$ and then compute for each vertex $i$ its level in the tree (distance from the root); we assign the color $1$ to the vertices on even levels and the color $2$ to those on odd levels. However, in some situations, we might be forced to process the vertices in a given order. In this case, it would be useful to compute the worst-case coloring that can be obtained by this heuristic, i.e. the largest number of colors that are used, under the worst-case ordering of the tree vertices ({\it Grundy number}). I will present an $O(n \cdot log(log(n)))$ algorithm for this problem, similar in nature to the linear algorithm presented in [4]. For each vertex $i$, we will compute $cmax(i)$=the largest color the can be assigned to vertex $i$ in the worst-case, if vertex $i$ is the last vertex to be colored. The value $max\{cmax(i)|1\le i\le n\}$ is the largest number of colors that can be assigned by the first fit online coloring heuristic.

We will root the tree at an arbitrary vertex $r$. The algorithm consists of two stages. In the first stage, the tree is traversed bottom-up and for each vertex $i$ we compute $c(1,i)$=the largest color that can be assigned to vertex $i$, considering only the tree $T(i)$. For a leaf vertex $i$, we have $c(1,i)=1$. For a non-leaf vertex $i$, we will sort its sons $s(i,1), \ldots, s(i,ns(i))$, such that $c(1,s(i,1))\le c(1,s(i,2))\le \ldots \le c(1,s(i,ns(i)))$. We will initialize $c(1,i)$ to $1$ and then consider the sons in the sorted order. When we reach son $s(i,j)$, we compare $c(1,s(i,j))$ with $c(1,i)$. If $c(1,s(i,j)) \ge c(1,i)$, then we increment $c(1,i)$ by $1$ (otherwise, $c(1,i)$ stays the same). The justification of this algorithm is the following: if a vertex $i$ can be assigned color $c(1,i)$ in some ordering of the vertices in $T(i)$, then there exists an ordering in which it can be assigned any other color $c'$, such that $1\le c'\le c(1,i)$. Then, when traversing the sons and reaching a son $s(i,j)$ with $c(1,s(i,j)) \ge c(1,i)$, we consider an ordering of the vertices in $T(s(i,j))$, where the color of vertex $s(i,j)$ is $c(1,i)$; thus, we can increase the maximum color that can be assigned to vertex $i$.

After the bottom-up tree traversal, we have $cmax(r)=c(1,r)$, but we still have to compute the values $cmax(i)$ for the other vertices of the tree. We could do that by rooting the tree at every vertex $i$ and running the previously described algorithm, but this would take $O(n^2\cdot log(log(n)))$ time. However, we can compute these values faster, by traversing the tree vertices in a top-down manner (considering the tree rooted at $r$). For each vertex $i$, we will compute $colmax(parent(i),i)$=the maximum color that can be assigned to $parent(i)$ if we remove $T(i)$ from the tree and afterwards we consider $parent(i)$ to be the (new) root of the tree. We will use the values $c(2,i)$ as temporary storage variables. $c(2,i)$ is initialized to $c(1,i)$, for every vertex $i$. When computing $cmax(i)$, we consider that vertex $i$ is the root of the tree. Let's assume that we computed the value $cmax(i)$ of a vertex $i$ and now we want to compute the value $cmax(j)$ of a vertex $j$ which is a son of vertex $i$. We remove $j$ from the list of sons of vertex $i$ and add $parent(i)$ to this list ($parent(i)$=vertex $i$'s parent in the tree rooted at the initial vertex $r$). We now need to lift vertex $j$ above vertex $i$ and make $j$ the new root of the tree. In order to do this, we will recompute the value $c(2,i)$, which is computed similarly to $c(1,i)$, except that we consider the new list of sons for vertex $i$ (and their $c(2,*)$ values). Afterwards, we add vertex $i$ to the list of sons of vertex $j$. We will compute the value $cmax(j)$ similarly to the value $c(1,j)$, using the values $c(2,*)$ of vertex $j$'s sons (instead of the $c(1,*)$ values of the sons). After computing $cmax(j)$ we restore the lists of sons of vertices $i$ and $j$ to their original states (as if the tree were rooted at the initial vertex $r$). After computing the values $cmax(u)$ of all the descendants $u$ of a vertex $j$, we reset the value $c(2,j)$ to $c(1,j)$.
 
Both traversals take $O(n\cdot log(n))$ time, if we sort the $ns(i)$ sons of every vertex $i$ in $O(ns(i)\cdot log(ns(i)))$ time. However, it has been proved in [4] that the minimum number of vertices of a tree with the Grundy number $q$ is $2^{q-1}$, which is the binomial tree $B(q-1)$. The binomial tree $B(0)$ consists of only one vertex. The binomial tree $B(k \ge 1)$ has a root vertex with $k$ neighbors; the $i^{th}$ of these neighbors $(0\le i\le k-1)$ is the root of a $B(i)$ binomial tree. Thus, every value $c(1,*)$, $c(2,*)$ and $cmax(*)$ can be represented using $O(log(log(n)))$ bits.  We can use radix-sort and obtain an $O(n\cdot log(log(n)))$ time complexity. The pseudocode of the functions is given below. The main algorithm consists of calling {\it FirstFit-BottomUp(r)}, initializing the $c(2,*)$ values to the $c(1,*)$ values, setting $cmax(r)=c(1,r)$ and then calling {\it FirstFit-TopDown(r)}\vspace{2mm}

\noindent \hspace{2mm}
\underline{\bf Compute(i, idx):} {\it

\noindent \hspace{2mm}
{\bf sort} the sons of vertex i, such that c(idx,s(i,1))$\le$ \ldots $\le$c(idx,s(i,ns(i)))

\noindent \hspace{2mm}
c(idx,i)=1

\noindent \hspace{2mm}
{\bf for} j=1 {\bf to} ns(i) {\bf do} {\bf if} (c(idx,s(i,j))$\ge$c(idx,i)) {\bf then} c(idx,i)=c(idx,i)+1
} \vspace{2mm}

\noindent \hspace{2mm}
\underline{\bf FirstFit-BottomUp(i):} {\it

\noindent \hspace{2mm}
{\bf for} j=1 {\bf to} ns(i) {\bf do} {\bf FirstFit-BottomUp}(s(i,j))

\noindent \hspace{2mm}
{\bf Compute}(i, 1)
} \vspace{2mm}

\noindent \hspace{2mm}
\underline{\bf FirstFit-TopDown(i):} {\it

\noindent \hspace{2mm}
{\bf if} $(i \neq r)$ {\bf then}

\noindent \hspace{4mm}
    {\bf remove} vertex i from the list of sons of parent(i)

\noindent \hspace{4mm}
    {\bf add} parent(parent(i)) to the list of sons of parent(i) (if parent(i) $\neq$ r)

\noindent \hspace{4mm}
    {\bf Compute}(parent(i),2); colmax(parent(i),i)=c(2,parent(i))

\noindent \hspace{4mm}
    {\bf add} parent(i) to the list of sons of vertex i

\noindent \hspace{4mm}
    {\bf Compute}(i,2); cmax(i)=c(2,i)

\noindent \hspace{4mm}
    {\bf restore} the original lists of sons of the vertices parent(i) and i
  
\noindent \hspace{2mm}
{\bf for} j=1 {\bf to} ns(i) {\bf do} {\bf FirstFit-TopDown}(s(i,j))

\noindent \hspace{2mm}
c(2,i)=c(1,i)
}

\section{Other Optimization and Counting Problems}

\subsection{Building a (Constrained) Tree with Minimum Height}

In this subsection I consider the following optimization problem: We are given a sequence of $n$ leaves and each leaf $i$ $(1\le i\le n)$ has a height $h(i)$. We want to construct a (strict) binary tree with $n-1$ internal nodes, such that, in an inorder traversal of the tree, we encounter the $n$ leaves in the given order. The height of an internal node $i$ is $h(i)=1+max\{h(leftson(i), h(rightson(i))\}$ (the height of the leaves is given). We are interested in computing a tree whose root has minimum height. A straight-forward dynamic programming solution is the following: compute $Hmin(i,j)$=the minimum height of a tree containing the leaves $i$, $i+1$, \ldots, $j$. We have: $Hmin(i,j)$=$1 + min_{i\le k\le j-1}$ $max\{Hmin(i,k),$ $Hmin(k+1,j)\}$. $Hmin(1,n)$ is the answer to our problem. However, the time complexity of this algorithm is $O(n^3)$, which is unsatisfactory. An optimal, linear-time algorithm was given in [14]. The main idea of this algorithm is the following. We traverse the leaves from left to right and maintain information about the rightmost path of the optimal tree for the first $i$ leaves. Then, we can add the $(i+1)^{st}$ leaf by modifying the rightmost path of the optimal tree for the first $i$ leaves. Let's assume that we processed the first $i$ leaves and the optimal tree for these leaves contains, on its rightmost path, the vertices $v(1)$, $v(2)$, \ldots, $v(nv(i))$, in order, from the root to the rightmost leaf ($v(1)$ is the root). Let's assume that the heights of the subtrees rooted at these vertices are $hv(1)$, \ldots, $hv(nv(i))$. It is easy to build this tree for $i=1$ and $i=2$ (it is unique). When adding the $(i+1)^{st}$ leaf, we traverse the rightmost path from $nv(i)$ down to 2. Assume that we are considering the vertex $v(j)$. If $hv(j-1) < (2+max\{hv(j), h(i+1)\})$, then we disconsider the vertex $v(j)$ from the rightmost path and move to the next vertex $(v(j-1))$. Let's assume that the path now contains the vertices $v(1)$, \ldots, $v(nv'(i))$. We replace vertex $v(nv'(i))$ by a new vertex $vnew$, whose left son will be $v(nv'(i))$ (together with its subtree) and whose right son will be the $(i+1)^{st}$ leaf. The height of the new vertex will be $1+max\{hv(nv'(i)), h(i+1)\}$. The rightmost path of the optimal tree behaves like a stack and, thus, the overall time complexity is linear.

I will present a sub-optimal $O(n\cdot log(n))$ time algorithm which is interesting on its own. The algorithm is similar to Huffman's algorithm for computing optimal prefix-free codes, except that it maintains the order of the leaves. A suggestion that such an approach might work was given to me by C. Gheorghe. At step $i$ $(1\le i\le n-1)$ of the algorithm, we will have $n-i+1$ subtrees of the optimal tree. Each subtree $j$ contains an interval of leaves $[leftleaf(j),rightleaf(j)]$ and its height is $h(j)$. We will combine the two adjacent subtrees $j$ and $j+1$ whose combined height {\it (1+max\{height(subtree j),height(subtree j+1)\})} is minimum among all the $O(n)$ pairs of adjacent subtrees. At the first step, the $n$ subtrees are represented by the $n$ leaves, whose heights are given. A straight-forward implementation of this idea leads to an $O(n^2)$ algorithm. However, the processing time can be improved by using two segment trees [15], $A$ and $B$, with $n$ and $n-1$ leaves, respectively. Each node $q$ of a segment tree corresponds to an interval of leaves $[left(q), right(q)]$ (leaves are numbered starting from $1$). Each leaf node of the segment tree $A$ can be in the {\it active} or {\it inactive} state. Each node $q$ of $A$ (whether leaf or internal node) maintains a value $nactive(q)$, denoting the number of active leaves in its subtree. Initially, each of the $n$ leaves of $A$ is active and the $nactive(*)$ values are initialized appropriately, in a bottom-up manner ($1$, for a leaf node, and $nactive(leftson(q)) + nactive(rightson(q))$, for an internal node $q$). Segment tree $B$ has $n-1$ leaves and each node of $B$ (leaf or internal node) stores a value $hc$. If leaf $i$ $(1\le i\le n-1)$ is {\it active} in $A$, then {\it hc(leaf i)=1+max\{h(i), h(j)\}}, where $j>i$ is the next active leaf. If leaf $i$ is not {\it active} in $A$ or is the last {\it active} leaf, then {\it hc(leaf i)=$+\infty$}. The value $hc$ of each internal node $q$ of $B$ is the minimum among all the $hc$ values of the leaves in node $q$'s subtree, i.e. {\it hc(node q)=min\{hc(leftson(q)), hc(rightson(q))\}}. Moreover, each node $q$ of $B$ maintains the number $lnum$ of the leaf in its subtree which gives the value {\it hc(node q)}. We have {\it lnum(leaf i)=i} and {\it lnum(internal node q)=if (hc(leftson(q)) $\le$ hc(rightson(q))) then lnum(leftson(q)) else lnum(rightson(q))}.

At each step $i$ $(1\le i\le n-1)$, each {\it active} leaf is the leftmost leaf of a subtree of the optimal tree. After every step, the number of active leaves decreases by $1$. We can find in $O(log(n))$ time the pair of adjacent subtrees to combine. The height of the combination of these subtrees is {\it hc(root node of B)}, the leftmost leaf of the first subtree is {\it i=lnum(root node of B)} and that of the second subtree is $j=next\_active(i)$. We define the function $next\_active$ by using two other functions: $rank(i)$ and $unrank(r)$. $rank(i)$ returns the number of {\it active} leaves before leaf $i$ ($0\le rank(i)\le$ {\it nactive(root node of A)-1}). $unrank(r)$ returns the index of the leaf whose rank is $r$. The two functions are inverses of each other: $unrank(rank(i))=i$ and $rank(unrank(r))=r$. We have {\it rank(i)=rank'(i, root node of A)}, {\it unrank(r)=unrank'(r, root node of A)} and {\it next\_active(i)=unrank(rank(i) + 1)}.\vspace{2mm}

\noindent \hspace{2mm}
\underline{\bf rank'(i, q):} {\it

\noindent \hspace{2mm}
{\bf if} (q is a leaf node) {\bf then}

\noindent \hspace{4mm} 
  {\bf if} (left(q)=right(q)=i) {\bf then return}(0) {\bf else return}(-1)

\noindent \hspace{2mm}
{\bf else} {\bf if} ($i>right(leftson(q))$) {\bf then} 

\noindent \hspace{4mm}
{\bf return}(nactive(leftson(q))+rank'(i, rightson(q)))

\noindent \hspace{2mm}
{\bf else return}(rank'(i, leftson(q)))
} \vspace{2mm}

\noindent \hspace{2mm}
\underline{\bf unrank'(r, q):} {\it

\noindent \hspace{2mm}
{\bf if} (q is a leaf node) {\bf then}

\noindent \hspace{4mm}
  {\bf if} $(r>0)$ {\bf then return}(-1) {\bf else return}(left(q))

\noindent \hspace{2mm}
{\bf else} {\bf if} $(nactive(leftson(q))\le r)$ {\bf then} 

\noindent \hspace{4mm}
{\bf return}(unrank'(r-nactive(leftson(q)), rightson(q)))

\noindent \hspace{2mm}
  {\bf else return}(unrank'(r, leftson(q)))
} \vspace{2mm}

The functions {\it rank}, {\it unrank} and {\it next\_active} take $O(log(n))$ time each. After obtaining the indices of the two active leaves $i$ and $j$ whose corresponding subtrees are united (by adding a new internal node whose left son is the root of $i$'s subtree and whose right son is the root of $j$'s subtree), we mark leaf $j$ as {\it inactive}. We do this by traversing the segment tree $A$ from leaf $j$ towards the root (from $j$ to $parent(j)$, $parent(parent(j))$, \ldots, {\it root node of A}) and decrement by $1$ the $nactive$ values of the visited nodes. Then, we change the $h$ values of leaves $i$ and $j$. We set {\it h(i)=hc(root node of B)} and $h(j)=+\infty$. After this, we will also change the $hc$ values associated to the leaves $i$ and $j$ in the segment tree $B$. The new $hc$ value of leaf $j$ will be $+\infty$. If $i$ is now the last active leaf, then {\it hc(leaf i)} becomes $+\infty$, too. Otherwise, let $j'=next\_active(i)$, the next {\it active} leaf after $i$ (at this point, leaf $j$ is not {\it active} anymore). We will change {\it hc(leaf node i)} to $(1+max\{h(i), h(j')\})$. After changing the $hc$ value of a leaf $k$, we traverse the tree from leaf $k$ towards the root (visiting all of $k$'s ancestors, in order, starting from $parent(k)$ and ending at the root of $B$). For each ancestor node $q$, we recompute {\it hc(node q)} as $min\{hc(leftson(q)), hc(rightson(q))\}$.

\subsection{The Number of Trees with a Fixed Number of Leaves}

In order to compute the number of labeled trees with $n$ vertices and exactly $p$ leaves, we will compute a table $NT(i,j)$=the number of trees with $i$ vertices and exactly $j$ leaves $(1\le j\le i\le n)$. Obviously, we have $NT(1,1)=NT(2,2)=1$ and $NT(i,j)=0$ for $i=1,2$ and $j\neq i$. For $i > 2$, we have $NT(i,i)=0$ and for $1\le j\le i-1$, we will proceed as follows. The $j$ leaves can be chosen in $C(i,j)$ ways ({\it $i$ choose $j$}). After choosing the identifiers of the $j$ leaves, we will conceptually remove the leaves from the tree, thus remaining with a tree having $i-j$ vertices and any number of leaves $k$ $(1\le k\le j)$. Each of the $j$ leaves that we conceptually removed is adjacent to one of these $k$ vertices. Furthermore, each of these $k$ vertices is adjacent to at least one of the $j$ leaves from the larger tree. Thus, we need to compute the number of surjective functions $f$ from a domain of size $j$ to a domain of size $k$. We will denote this value by $NF(j,k)$. This is a "classical" problem, but I will present a simple solution, nevertheless. We have $NF(0,0)=1$ and $NF(j,k)=0$, if $j < k$. In order to compute the values for $k\ge 1$ and $j\ge k$, we will consider every number $g$ of values $x$ from the set $\{1,\ldots,j\}$ for which $f(x)=k$. Once $g$ is fixed, we have $C(j,g)$ ways of choosing the $g$ values from the set $\{1,\ldots,j\}$. For each such possibility we have $NF(j-g, k-1)$ ways of extending it to a surjective function. Thus, $NF(j,k)=\sum_{g=1}^{j} C(j,g)\cdot NF(j-g, k-1)$. We can tabulate all the $NF(*,*)$ values in $O(n^3)$ time (after tabulating the combinations $C(*,*)$ in $O(n^2)$ time, first). With the $NF(*,*)$ values computed, we have $NT(i,j)=C(i,j) \cdot  \sum_{k=1}^{j}(NT(i-j,k)\cdot NF(j,k))$. We can easily compute each entry $NT(i,j)$ in $O(n)$ time, obtaining an $O(n^3)$ overall time complexity. The technique of performing dynamic programming on successive layers of leaves of a tree is also useful in several other counting problems.

\subsection{The Number of Trees with Degree Constraints}

We want to compute the number of unlabeled, rooted trees with $n\ge 2$ vertices, such that the (degree / number of sons) of each vertex belongs to a set $S$, which is a subset of $\{0,1,2,\ldots,n-1\}$. By {\it(a/b)} we mean that $a$ refers to the degree-constrained problem and $b$ refers to the number-of-sons-constrained problem (everything else being the same). Because every tree with $n\ge 2$ vertices must contain at least a leaf (a vertex of degree $1$) and at least one vertex with at least $1$ son, the set $S$ will always contain the subset ($\{1\}/\{0,1\}$). We will compute a table $NT(i, j, p)$=the number of trees with $i$ vertices, such that the root has degree $j$ ($j$ sons) and the maximum number of vertices in the subtree of any son of the root is $p$; moreover, except perhaps the tree root, the (degrees/numbers of sons) of all the other vertices belong to the set $S$. Because the trees are unlabeled, we can sort the sons of each vertex in non-decreasing order of the numbers of vertices in their subtrees. Thus, we will compute the table $NT$ in increasing order of $p$. $NT(1, 0, p)=1$ and $NT(1, j>0, p)=NT(i \ge 2, j, 0)=0$. For $p\ge 1$ and $i\ge 2$, we have:

$NT(i, j, p)$=$NT(i, j, p-1)$+$\sum_{k=1}^{\left\lfloor\frac{i-1}{p}\right\rfloor}$ $NT(i-k\cdot p, j-k, p-1)\cdot $ $CR(TT(p), k)$

$TT(p)$ is the total number of trees with $p$ vertices, for which the (degree / number of sons) of the root is equal to some ($(x-1)$/$(x)$), $x \in S$, and the (degrees / numbers of sons) of the other vertices belong to the set $S$. By $CR(i,j)$ we denote combinations with repetitions of $i$ elements, out of which we choose $j$. Because the argument $i$ can be very large, we cannot tabulate $CR(i, j)$. Instead, we will compute it on the fly. We know that $CR(i, j)=C(i+j-1, j)$ and that $C(i, j)= \frac{i-j+1}{j} \cdot C(i, j-1)$. Thus, $CR(i,j)$ can be computed in $O(j)$ time. Before computing any value $NT(*,*,p)$, we need to compute and store the values $TT(p)$, $TT(p)=\sum_{x \in S} NT(p, ((x-1)/(x)), p-1)$, and $CR(TT(p), k)$, for all the values of $k$ $(1\le k\le \left\lfloor \frac{n-1}{p} \right\rfloor)$. We can compute all of these values in $O(n^3\cdot log(n))$ time. The desired number of trees is $\sum_{x \in S} NT(n, x, n-1)$. The memory storage can be reduced from $O(n^3)$ to $O(n^2)$, by noticing that the values $NT(*,*,p)$ are computed based only on the values $NT(*,*,p-1)$. Thus, we can maintain these values only for the most recent two values of $p$.

A less efficient method is to compute the numbers $Tok(i)$=the number of trees with $i$ vertices, such that each vertex satisfies the (degree/number of sons) constraints. $Tok(1)=Tok(2)=1$. We will make use of the $TT(i)$ values defined previously, except that they will be computed differently. For every $i \ge 2$, we consider every possible number $x$ of sons of the tree root and compute $NT2(i,x)$=the number of trees with $i$ vertices, such that the tree root has $x$ sons and all the other vertices satisfy the (degree/number of sons) constraints. We will generate all the possibilities $(y(1), y(2), ..., y(i-1))$, with $0 \le y(j) \le \left\lfloor \frac{i-1}{j} \right\rfloor$ $(1 \le j \le i-1)$ and $y(1)$ + \ldots + $y(i-1)$=$x$. $y(j)$ is the number of sons of the tree root which have $j$ vertices in their subtrees. The number of trees "matching" such a partition is equal to $\prod_{j=1}^{i-1}CR(TT(j), y(j))$. $NT2(i,x)$ is computed by summing the numbers of trees "matching" every partition. Afterwards, if $x \in S$, we add $NT2(i,x)$ to $Tok(i)$. If $x=((y-1)/(y))$ and $y \in S$, then we add $NT2(i,x)$ to $TT(i)$. $NT2(i,x)$ may be added to both $Tok(i)$ and $TT(i)$.

\section{Related Work}

Reliability analysis and improvement techniques for distributed systems were considered in [6,7]. Reliability analysis and optimization for tree networks in particular were considered in [3,5,8]. Different kinds of tree partitioning algorithms, based on optimizing several objectives, were proposed in [9,10,16]. Problems related to tree coloring were studied in [4]. Content delivery in distributed systems is a subject of high practical and theoretical interest and is studied from multiple perspectives. Communication scheduling in tree networks was considered in many papers (e.g. [17]) and the optimization of content delivery trees (multicast trees) was studied in [11].

\section{Conclusions and Future Work}

In this paper I considered several optimization problems regarding distributed systems with tree topologies (e.g. peer-to-peer networks, wireless networks, Grids), which have many practical applications: minimum weight cycle completion (reliability improvement), constrained partitioning (distributed coordination and control), minimum number of streams and degree-constrained minimum spanning trees (efficient content delivery), optimal matchings (data replication and resource allocation), coloring (resource management and frequency allocation) and tree counting aspects. All these problems are variations or extensions of problems which have been previously posed in other research papers. The presented techniques are either better (faster or more general) than the previous solutions or easier to implement.

\vspace{3mm}

\noindent
{\small {\bf References}

\noindent
1. J. Roskind, R. E. Tarjan, {\it A Note on Finding Minimum-Cost Edge-Disjoint Spanning Trees}, Mathematics and Operations Research {\bf 10 (4)} (1985), 701-708.

\noindent
2. M. A. Bender, M. Farach-Colton, {\it The LCA Problem revisited},  Lecture Notes in Computer Science {\bf 1776} (2000), 88-94.

\noindent
3. M. Scortaru, {\it National Olympiad in Informatics}, Gazeta de \
informatica (Informatics Gazzette) {\bf 12 (7)} (2002), 8-13.

\noindent
4. S. M. Hedetniemi, S. T. Hedetniemi, T. Beyer, {\it A Linear Algorithm for the Grundy (Coloring) Number of a Tree}, Congressus Numerantium {\bf 36} (1982), 351-362.

\noindent
5. M. I. Andreica, N. Tapus, {\it Reliability Analysis of Tree Networks Applied to Balanced Content Replication}, Proc. of the IEEE Intl. Conf. on Automation, Robotics, Quality and Testing (2008), 79-84.

\noindent
6. D. J. Chen, T. H. Huang, {\it Reliability Analysis of Distributed Systems Based on a Fast Reliability Algorithm}, IEEE Trans. on Par. and Dist. Syst. {\bf 3} (1992), 139-154.

\noindent
7. A. Kumar, A. S. Elmaghraby, S. P. Ahuja, {\it Performance and reliability optimization for distributed computing systems}, Proc. of the IEEE Symp. on Comp. and Comm. (1998), 611-615.

\noindent
8. H. Abachi, A.-J. Walker, {\it Reliability analysis of tree, torus and hypercube message passing architectures}, Proc. of the IEEE S.-E. Symp. on System Theory (1997), 44-48.

\noindent
9. G. N. Frederickson, {\it Optimal algorithms for tree partitioning}, Proc. of the ACM-SIAM Symposium on Discrete Algorithms (SODA) (1991), 168-177.

\noindent
10. R. Cordone, {\it A subexponential algorithm for the coloured tree partition problem}, Discrete Applied Mathematics {\bf 155 (10)} (2007), 1326-1335.

\noindent
11. Y. Cui, Y. Xue, K. Nahrstedt, {\it Maxmin overlay multicast: rate allocation and tree construction}, Proc. of the IEEE Workshop on QoS (IWQOS) (2004), 221-231. 

\noindent
12. Y. Qinglin, {\it Factors and Factor Extensions}, M.Sc. Thesis, Shandong Univ., 1985.

\noindent
13. T. L. Magnanti, L. A. Wolsey, {\it Optimal Trees}, Handbooks in Operations Research and Management Science, {\bf vol. 7, chap. 9} (1995), 513-616.

\noindent
14. S.-C. Mu, R. S. Bird, {\it On Building Trees with Minimum Height, Relationally}, Proc. of the Asian Workshop on Programming Languages and Systems (2000).

\noindent
15. M. I. Andreica, N. Tapus, {\it Optimal Offline TCP Sender Buffer Management Strategy}, Proc. of the Intl. Conf. on Comm. Theory, Reliab., and QoS (2008), 41-46.

\noindent
16. B. Y. Wu, H.-L. Wang, S. T. Kuan, K.-M. Chao, {\it On the Uniform Edge-Partition of a Tree}, Discrete Applied Mathematics {\bf 155 (10)} (2007), 1213-1223.

\noindent
17. M. R. Henzinger, S. Leonardi, {\it Scheduling multicasts on unit-capacity trees and meshes}, J. of Comp. and Syst. Sci. {\bf 66 (3)} (2003), 567-611.

\noindent
18. T. H. Cormen, C. E. Leiserson, R. L. Rivest, C. Stein, {\it Introduction to Algorithms}, MIT Press and McGraw-Hill (2001).

\end{document}